\documentclass[reprint,aps,superscriptaddress,prl]{revtex4-2}
\usepackage{physics}
\usepackage{graphicx}
\usepackage{amssymb}
\usepackage{amsthm}
\usepackage{bm}
\usepackage{txfonts}
\usepackage{pstricks}
\usepackage[colorlinks=true,allcolors=blue]{hyperref} 
\usepackage[capitalise]{cleveref}

\newcommand{\vk}{{\bm{k}}}
\definecolor{mygreen}{rgb}{0.15, 0.6, 0.15}
\definecolor{mygrey}{rgb}{0.5, 0.5, 0.5}

\begin{document}
\title{Ground-state properties of metallic solids from \textit{ab initio} coupled-cluster theory}
\author{Verena A. Neufeld}
\affiliation{Department of Chemistry, Columbia University, New York, New York 10027, USA}
\author{Hong-Zhou Ye}
\affiliation{Department of Chemistry, Columbia University, New York, New York 10027, USA}
\author{Timothy C. Berkelbach}
\affiliation{Department of Chemistry, Columbia University, New York, New York 10027, USA}
\affiliation{Center for Computational Quantum Physics, Flatiron Institute, New York, New York 10010, USA}
\date{\today}  

\begin{abstract}
Metallic solids are a challenging target for wavefunction-based electronic structure
theories and have not been studied in great detail by such methods.
Here, we use coupled-cluster theory with single and double excitations (CCSD) to study
the structure of solid lithium and aluminum using optimized Gaussian basis sets. 
We calculate the equilibrium lattice
constant, bulk modulus, and cohesive energy and compare them to experimental
values, finding accuracy comparable to common density functionals.
Because the quantum chemical ``gold standard'' CCSD(T) (CCSD with perturbative
triple excitations) is inapplicable to metals in the thermodynamic limit, we
test two approximate improvements to CCSD, which are found to improve the
predicted cohesive energies.
\end{abstract}

\maketitle

\textit{Introduction.}
Ab initio wavefunction-based electronic structure theories are being
increasingly applied to periodic solids~\cite{hirata_coupled-cluster_2004,
marsman_second-order_2009, gruneis_second-order_2010, booth_towards_2013,
liao_communication_2016, mcclain_gaussian-based_2017, gruber_applying_2018,
hummel_finite_2018, liao_comparative_2019, azadi_equation_2020,
gao_electronic_2020, lange_active_2020, pulkin_first-principles_2020}, where
they can be used as predictive tools on their own or to guide the choice of
functionals in more affordable density functional theory (DFT)~\cite{hohenberg_inhomogeneous_1964, kohn_self-consistent_1965} calculations.  To
date, most applications of these methods are to semiconducting or insulating
systems.  Extending to metals is a challenge because many of the most successful
wavefunction-based methods employ finite-order perturbation theory, whose
correlation energy typically diverges in the thermodynamic
limit~\cite{macke_uber_1950,gruneis_second-order_2010, shepherd_many-body_2013,callahan_dynamical_2021}.  Coupled-cluster (CC)
theory~\cite{bartlett_coupled-cluster_2007} is particularly promising in
this regard, because even its lowest-order nontrivial truncation to single and
double excitations (CCSD) includes a number of important, canonical classes of
diagrams, including ladder diagrams (important at low density) and ring diagrams
(important at high density and necessary to remove the aforementioned
divergence)~\cite{gell-mann_correlation_1957,bishop_electron_1978,bishop_electron_1982,shepherd_range-separated_2014,irmler_duality_2019}.
Despite extensive application to the ground state of the uniform electron
gas (UEG)~\cite{freeman_coupled-cluster_1977,bishop_electron_1978,bishop_electron_1982,shepherd_convergence_2012, shepherd_many-body_2013,
shepherd_coupled_2014,shepherd_range-separated_2014, shepherd_communication_2016, neufeld_study_2017,
mihm_optimized_2019, liao_towards_2021, callahan_dynamical_2021}, CC theory has
seen limited application to atomistic
metals~\cite{stoll2009,voloshina_development_2010,hummel_finite_2018,mihm_effective_2021}.

Here, we apply CCSD to study the structural and energetic properties of two
simple metals, body-centered cubic (BCC) lithium and face-centered cubic (FCC)
aluminum.  We address two of the key technical hurdles associated especially
with the \textit{ab initio} study of metals, namely the removal of basis set error and
finite-size error. To address basis set error, we use system-specific
Gaussian-type orbital (GTO) basis sets~\cite{boys_electronic_1950,vandevondele_gaussian_2007} that are optimized to lower the total
energy and to lower the condition number of the overlap matrix.  We demonstrate
the success of this approach by comparing our results to those obtained with
plane-wave basis sets. To address finite-size error, we employ relatively dense
Brillouin zone samplings with twisted boundary conditions and subsequent
extrapolation. We then address the CCSD error by two approximate methods:
adding the correlation energy due to perturbative triples~\cite{raghavachari_fifth-order_1989} with a coarse Brillouin
zone sampling
or
scaling the CCSD correlation energy by a non-empirical factor determined by the UEG.

\textit{Methods.}
For BCC lithium, we use an isotropic primitive cell containing two atoms.
For FCC aluminum, we use two unit cells: an anisotropic primitive cell
containing two atoms and an isotropic cubic cell with four atoms.
Except where indicated, calculations were performed with the lattice constants
$a=3.5~\AA$ (Li) and $a=4.05~\AA$ (Al), which are close to the experimental values.
All calculations are performed at zero temperature using
PySCF~\cite{sun_pyscf_2018,sun_recent_2020,mcclain_gaussian-based_2017} with
libcint~\cite{sun_libcint_2015}, and GTO-based calculations were performed with
Gaussian density fitting~\cite{sun_gaussian_2017,ye_fast_2021,ye_tight_2021}.
We use GTH
pseudopotentials~\cite{goedecker_separable_1996,hartwigsen_relativistic_1998,cp2k_new_nodate,hutter_new_nodate},
correlating three electrons per atom, for both lithium and aluminum. 

The original GTO basis sets designed for use with GTH
pseudopotentials~\cite{vandevondele_gaussian_2007} were not optimized for
correlated calculations and furthermore only contain $s$, $p$ and $d$ functions. 
Therefore in this work, we re-optimize the GTH-DZVP (DZ), GTH-TZV2P (TZ), and
GTH-QZV3P (QZ) basis sets; for aluminum, we also added $f$ functions to the TZ
and QZ basis sets, which were found to be important in our testing.
In periodic solids, increasing the size of the basis set by brute force
frequently leads to linear dependencies, quantified by an overlap matrix with
large condition number, and concomitant numerical issues. Following similar
works~\cite{vandevondele_gaussian_2007,daga_gaussian_2020,morales_accelerating_2020,li_optimized_2021,zhou_material_2021},
here we optimize these basis functions for each solid by minimizing the cost
function
\begin{equation}
	\mathrm{cost} = E_\mathrm{HF} + E_\mathrm{c}^{(2)} + \gamma\ln(\mathrm{cond}(\mathbf{S}))
	\label{eq:cost}
\end{equation}
where $E_\mathrm{HF}$ is the Hartree-Fock (HF) energy, $E_\mathrm{c}^{(2)}$ is
the second-order M\o ller-Plesset perturbation theory
(MP2)~\cite{moller_note_1934} correlation energy, $\mathbf{S}$ is the periodic
overlap matrix of the GTO basis, and $\gamma = 10^{-4}~E_h$. For
this basis set optimization, we sampled the Brillouin zone with a uniform
mesh~\cite{monkhorst_special_1976} of $N_\vk=2^3$ $\vk$-points (Li) or
$N_\vk=1^3$ $\vk$-points (Al), including the $\Gamma$ point; with these boundary
conditions, the system is gapped and thus MP2 provides a well-defined and
computationally affordable correlation energy. At this level of theory, the
exponents and contraction coefficients of the GTO basis functions were optimized
in an approximately alternating fashion.  The optimization was started from the
original GTH basis set (with additional $f$ functions in TZ and QZ, for Al) and
the final basis functions may represent a local minimum of the cost function.
To avoid biasing the atomic structure, the cost function~(\ref{eq:cost}) was
averaged over three lattice parameters approximately spanning the range used in
later calculations. 
See the Supporting Materials (SM) for further details about 
pseudopotentials and our optimized basis sets. 

\begin{figure}
	\includegraphics[width=3.5in]{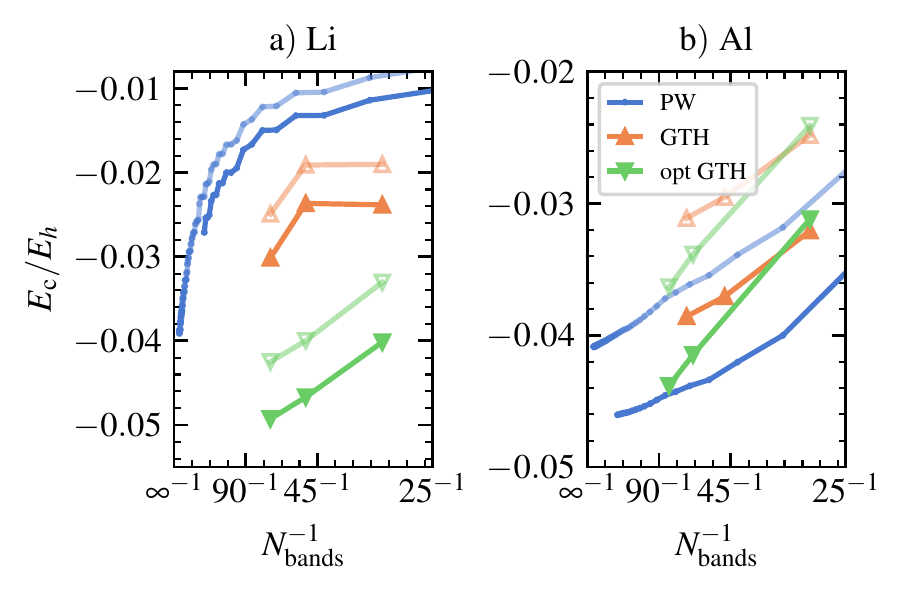}
	\caption{Basis set convergence of MP2 (faint, open symbols) and CCSD (solid, closed symbols) correlation energies $E_{\mathrm{c}}$ per atom
        for (a) lithium and (b) aluminum.
        Results were obtained using plane waves (PW) of increasing energy and using Gaussian type orbitals (GTO) and optimized GTOs, at the DZ, TZ, and QZ level.
        The Brillouin zone was sampled only at the $\Gamma$ point.
    }
	\label{fig:111_conv}
\end{figure}

To demonstrate the impact of basis set optimization, in Fig.~\ref{fig:111_conv} we show the basis set convergence of the MP2 and CCSD 
correlation energy of lithium and aluminum at the $\Gamma$ point of their primitive cells, 
comparing GTO and plane wave (PW) results
(for this figure only, the GTO results were evaluated using a PW basis set to compute
the occupied bands and an approximate PW resolution of the original and optimized GTO bases to 
compute the virtual 
bands~\cite{booth_plane_2016, morales_accelerating_2020,ye_correlation-consistent_2022}).
For Li, we see that GTO optimization increases
the correlation energy by almost a factor of two, i.e., about 20~m$E_h$. A crude extrapolation suggests that the optimized
QZ results recover about 80--90\% of the correlation energy in the basis set limit.
With this small $\vk$-point mesh,
we can perform PW calculations with a reasonably large number of orbitals, but for Li these results converge slowly due 
to the inclusion of the core $1s$ electrons. An MP2 calculation with our optimized QZ basis set with 67 bands recovers more correlation energy than one
with a PW basis set containing 1203 orbitals,
highlighting the immense computational savings afforded by GTO basis sets. With very large PW basis sets, we begin
to see the onset of $N_{\mathrm{bands}}^{-1}$ convergence to a basis set limit in good agreement with that of
the optimized GTO basis sets. For Al, the core electrons are not explicitly treated in the calculations and the PW 
calculations converge faster.
Again, we see the benefit of basis set optimization as well as the addition of $f$ functions.
The QZ result
captures about 90\% of the correlation energy in the basis set limit. 

In addition to recovering a greater amount of electron correlation, our basis sets were optimized to reduce their 
numerically problematic linear dependencies. Indeed, with our largest QZ basis, the condition number of 
the overlap matrix decreases from about $10^{13}$ to $10^4$ for Li and from about
$10^{16}$ to $10^{7}$ for Al.
For both Li and Al, the qualitative similarity between the MP2 and CCSD correlation
energies justifies our use of the former when optimizing the GTO basis set and suggests good transferability
to other correlated methods~\cite{morales_accelerating_2020}.
Henceforth, we use these optimized GTO basis sets. 
Testing (not shown) indicates that these system-specific optimized GTO basis sets 
perform similarly to the transferable,
correlation-consistent basis sets recently developed by two of us~\cite{ye_correlation-consistent_2022}.

\begin{figure}
	\includegraphics[width=3.5in]{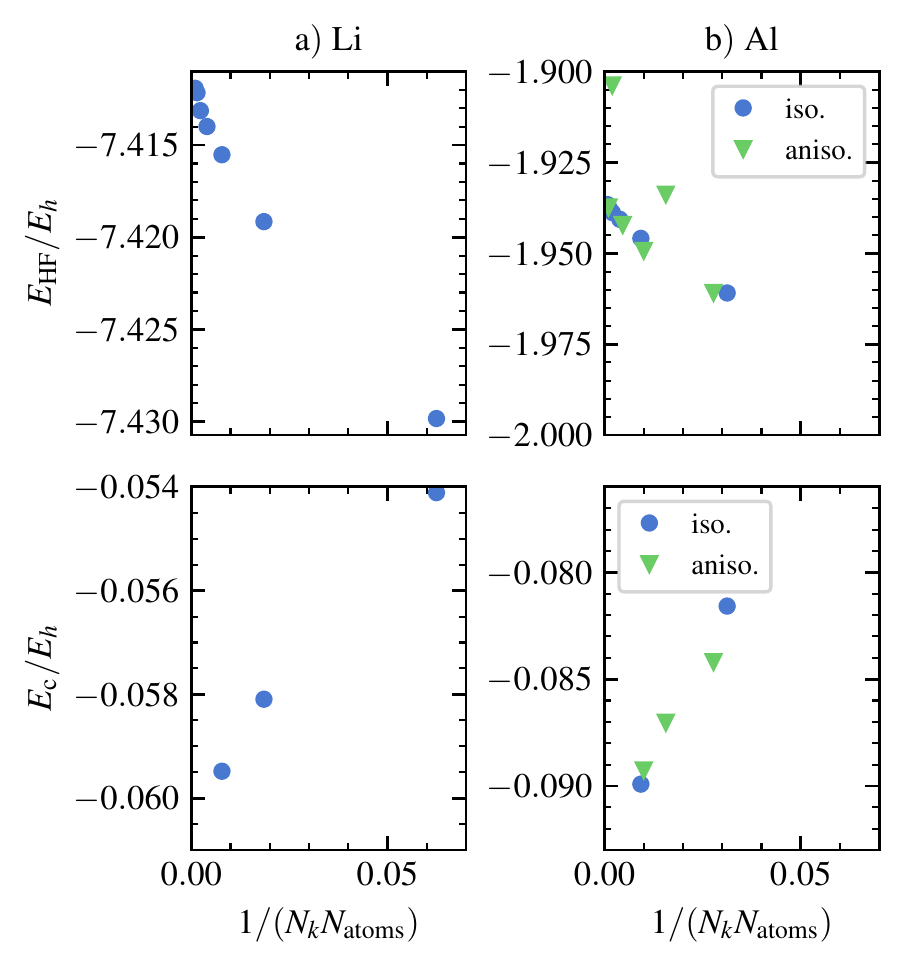}
	\caption{
               Thermodynamic
limit convergence of the DZ HF (top) and DZ CCSD correlation (bottom) energy per atom
as a function of the inverse of the number of atoms in the effective supercell
$N_\vk N_\mathrm{atoms}$, where $N_\vk$ is the number of $\vk$-points sampled in
the Brillouin zone and $N_\mathrm{atoms}$ is the number of atoms in the unit
cell. For Al, the results using an anisotropic primitive and an isotropic cubic cell are shown.
                }
	\label{fig:finite_size}
\end{figure}
Finite-size errors are especially problematic for metals
due to shell-filling effects, which can be alleviated with twisted
boundary conditions~\cite{baldereschi_mean-value_1973, chadi_special_1973, rajagopal_variational_1995,
lin_twist-averaged_2001, morris_hydrogensilicon_2008,mihm_optimized_2019, callahan_dynamical_2021, mihm_effective_2021}. 
In Fig.~\ref{fig:finite_size}, we show the finite size convergence of the HF energy and CCSD correlation energy
for Li and Al in the optimized DZ bases.
HF calculations were performed with up to 
$N_\vk=8^3$ (Li) and $N_\vk=7^3$ (Al) $\vk$-points, and a Madelung constant correction was used to eliminate the leading-order
$N_\vk^{-1/3}$ finite-size error~\cite{monkhorst_special_1976,paier_perdewburkeernzerhof_2005,sundararaman_regularization_2013} due to nonlocal exchange.
All calculations were performed using a twisted boundary condition defined by the Baldereschi 
point~\cite{baldereschi_mean-value_1973, rajagopal_variational_1995, morris_hydrogensilicon_2008}, 
which was found to yield smoother convergence to the thermodynamic limit (TDL)
than calculations without a twist angle.
Results for Li obtained by averaging over four Chadi-Cohen twist angles~\cite{chadi_special_1973}
(not shown) were found to give very similar results.

\begin{figure}[t]
\includegraphics[width=3.5in]{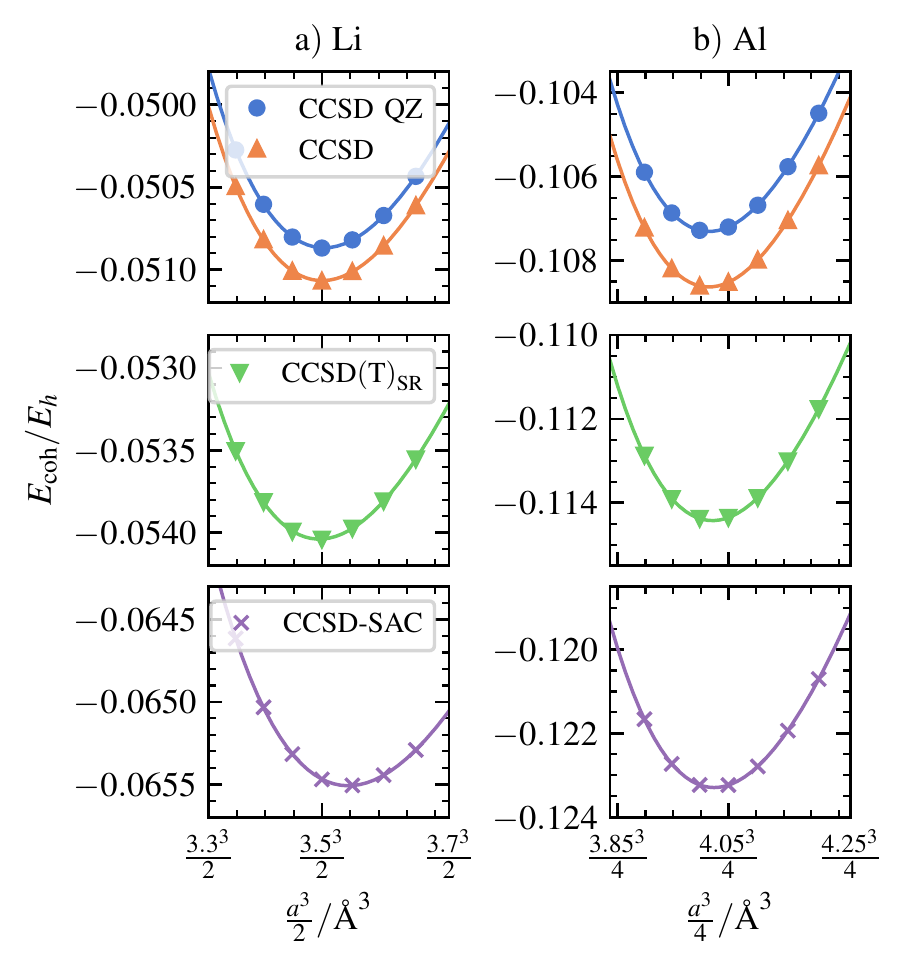}
	\caption{Cohesive energy equations of state from the indicated CCSD-based methods for a) 
        lithium and b) aluminum. The Birch--Murnaghan equation\cite{birch_finite_1947,zhang_performance_2018}
        was fit to the data and shown as a solid line.}
	\label{fig:binding}
\end{figure}

In this manner, the TDL of the HF energies can be estimated to an accuracy of about 1~m$E_h$.
For Li, CCSD calculations were performed with up to $N_\vk=4^3$ $\vk$-points;
with that mesh they were performed using a truncated basis of MP2 natural orbitals
and corrected based on results obtained with smaller meshes.
Extrapolation assuming finite-size errors that
scale as $N_\vk^{-1}$ suggests an extrapolation uncertainty of about 1~m$E_h$.
For Al, due to the larger number of atoms in a cubic unit cell, our largest 
mesh has $N_\vk=3^3$ $\vk$-points. 
For better extrapolation, we also performed calculations
using an anisotropic primitive cell, with up to $N_\vk=5\times 5\times 2$. 
Except for two meshes, all
results from both cell choices are found to lie roughly on a straight line and
reliable extrapolation can be performed
to estimate the HF energy in the TDL.
With the anisotropic cell, two meshes
($4\times 4\times 2$ and $8\times 8\times 4$) yield HF energies that are
too high, which may correspond to incorrect HF solutions due to the
challenge of minimization in metals at zero temperature; somewhat surprisingly,
the correlation energy associated with the higher-energy HF solution at
$N_\vk = 4\times 4\times 2$ is in line with the other correlation energies,
In all future calculations on Al, we use the isotropic cell for HF energies
and the anisotropic cell for correlation energies.
facilitating extrapolation.
The total energies that we calculate using our largest meshes differ 
from the extrapolated results by about 1~m$E_h$ (Li) and 3-4~m$E_h$ (Al),
and the extrapolated results have an uncertainty of about 1~m$E_h$ or less.

Henceforth, we assume basis set corrections and finite-size corrections are
independent and additive. To our DZ results obtained with increasingly large
$N_\vk$, we add basis set corrections determined by calculations with smaller
values of $N_\vk$. 
For the HF energy, we assume that the QZ result is near the complete basis set (CBS) limit. 
For the correlation energy, the CBS limit is estimated
via $X^{-3}$ extrapolation~\cite{helgaker_basis-set_1997} of
TZ and QZ results, where $X=3,4$ is the cardinality. 
See the SM for further details about our composite corrections.

\textit{Results.} In Fig.~\ref{fig:binding}, we show the equation of state (EOS) of Li
and Al as a function of the cell volume, 
where the cohesive energy, which is defined with respect to a single atom, was counterpoise-corrected
by surrounding the atom with ghost atoms and their basis 
functions~\cite{boys_calculation_1970,van_duijneveldt_state_1994,paulus_hartreefock_2007}. Single-atom calculations
were performed in a spin-unrestricted manner.
For Al, in sequential calculations from small to large volumes, we found that using 
the converged HF density matrix of the previous volume was essential in obtaining a smooth curve.
The HF EOS was calculated by performing TDL extrapolation at each lattice constant.
The correlation energy contribution to the CCSD EOS was calculated using 
our largest $\vk$-point meshes and then rigidly shifted by a finite-size correction
calculated at lattice parameters 3.5 \AA\ (Li) and 4.05 \AA\ (Al),
which are close to the experimental values. 

In Fig.~\ref{fig:binding}(a) and (b), we show CCSD results in the QZ basis and
CBS limit, which only differ by about 0.2~m$E_h$ for Li and 2~m$E_h$ for Al,
indicating the good performance of our optimized basis sets.  By fitting our
data to a Birch-Murnaghan equation~\cite{birch_finite_1947,zhang_performance_2018}, we
extract the lattice constant, bulk modulus, and cohesive energy. These
properties are listed in Tab.~\ref{tab:properties} and compared to experimental
values. The
experimental results have been corrected for zero-point motion (ZPM) using the
ZPM correction obtained in Ref.~\onlinecite{zhang_performance_2018} with
the HSE06 
functional~\cite{heyd_hybrid_2003,heyd_erratum_2006,krukau_influence_2006},
although other functionals yield similar corrections. 

\begin{table}[t]
\centering
\begin{tabular*}{0.48\textwidth}{@{\extracolsep{\fill}} llccc }
\hline\hline
&           & $a$ (\AA) & $B$ (GPa) & $E_\mathrm{coh}$ ($E_h$/atom) \\
\hline
Li &       &             &      &            \\
& HF (ours)&      3.68 & 9.0  & $-0.022$  \\
& HF~\cite{paulus_hartreefock_2007} &      3.73  & 8.1-11.9  & $-0.020$  \\
& CCSD (ours)&    3.50  & 12.5  & $-0.051$  \\
& Incremental CCSD~\cite{stoll2009}    &      3.50  & --  & $-0.060$  \\
& CCSD(T)$_\mathrm{SR}$  &      3.50  & 12.6  & $-0.054$  \\
& CCSD-SAC &      3.54  & 11.3  & $-0.066$  \\
& Exp.~\cite{felice_temperature_1977, berliner_effect_1986, kittel_intro_solid_2005, zhang_performance_2018} &    3.45  & 13.3  & $-0.061$ \\
&&&&\\
Al &       &             &      &            \\
& HF       &        4.08 & 80.0 & $-0.051$ \\
& CCSD     &      4.02  & 93.2  & $-0.109$  \\
& CCSD(T)$_\mathrm{SR}$  &      4.02  & 91.7  & $-0.114$  \\
& CCSD-SAC &      4.03  & 91.5  & $-0.123$  \\
& Exp.~\cite{kamm_lowtemperature_1964, giri_extrapolated_1985, kittel_intro_solid_2005, zhang_performance_2018}&    4.02  & 80.3 & $-0.126$ \\
\hline\hline
\end{tabular*}
\caption{
Lattice constant $a$, bulk modulus $B$, and cohesive energy $E_\mathrm{coh}$ of lithium 
and aluminum. All experimental (Exp.) values have been corrected for
zero-point motion (ZPM) effects as described in the text. The HF cohesive
energy of Li reported in Ref.~\onlinecite{paulus_hartreefock_2007} includes
a ZPM correction, which we estimate to be only 0.001~$E_h$.
}
\label{tab:properties}
\end{table}

\begin{figure*}[!]
	\includegraphics[width=6in]{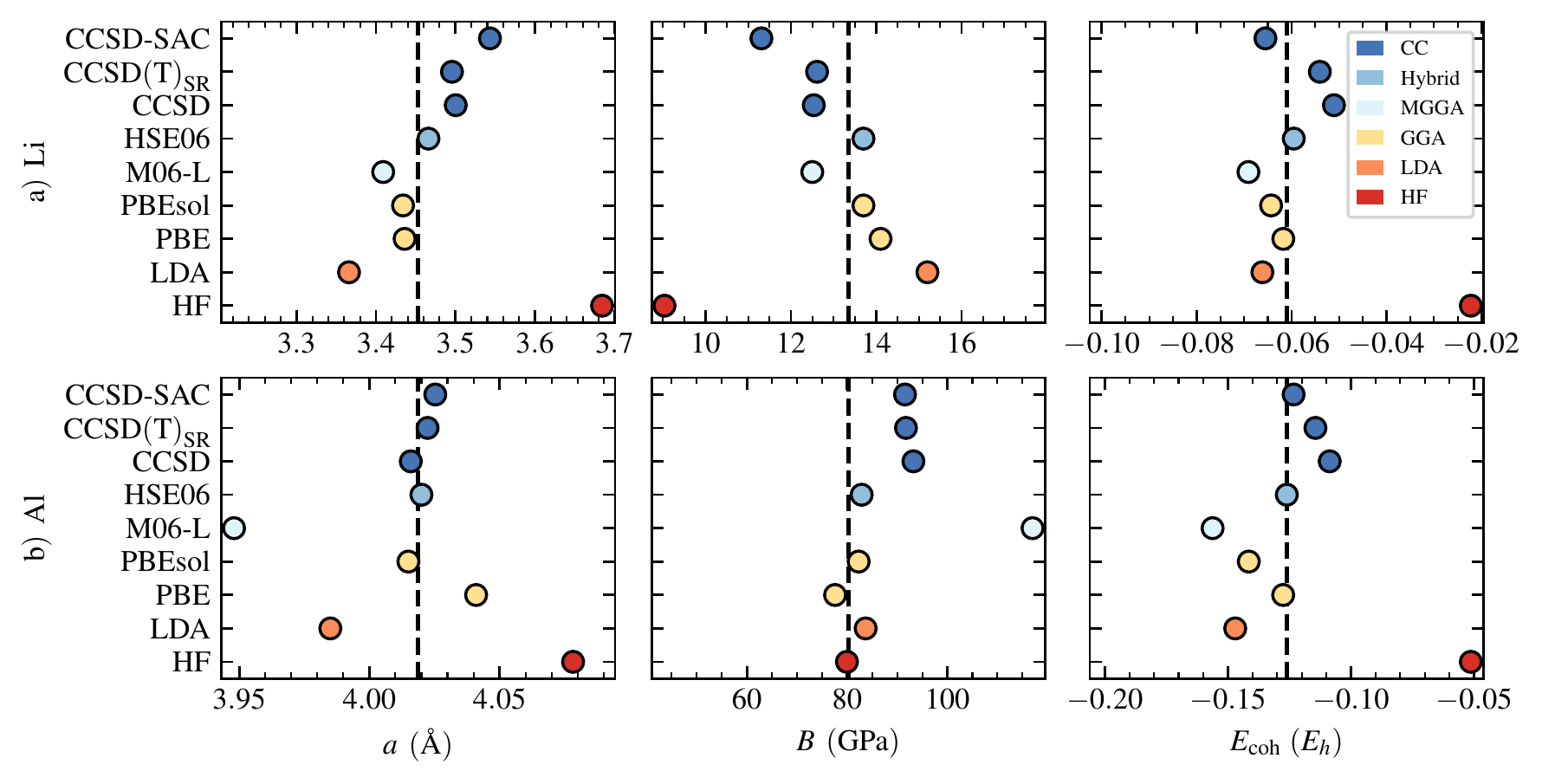}
        \caption{Structural properties (lattice parameter $a$, bulk modulus $B$,
and cohesive energy $E_{\mathrm{coh}}$) of a) lithium and b) aluminum evaluated
with HF (this work),
DFT (from Ref.~\onlinecite{zhang_performance_2018}) colored by rung, 
and coupled cluster (CC) (this work) in its CCSD, CCSD(T)$_\mathrm{SR}$
and UEG scaled CCSD (CCSD-SAC) forms. The zero or low temperature experimental
values (from Refs.~\onlinecite{kamm_lowtemperature_1964, giri_extrapolated_1985,
felice_temperature_1977, berliner_effect_1986, kittel_intro_solid_2005,
zhang_performance_2018}) are shown by dashed vertical lines. 
All experimental (Exp.) values have been corrected for
zero-point motion (ZPM) effects as described in the text.
}
	\label{fig:strucprop}
\end{figure*}

Comparing to experiment, we find that the magnitude of the CCSD cohesive energy
is too small by about 10~m$E_h$ (Li) and 20~m$E_h$ (Al), which our testing
suggests is mostly due to insufficient correlation in the solid rather than the
single atom. 
Therefore, we tested two approximate improvements to CCSD.  First, we tested
CCSD(T), which applies a perturbative correction to the correlation energy due
to triple excitations. The energy of the atom was calculated in the usual
manner, but the CCSD(T) energy of the solid must be calculated in a modified form
because otherwise it diverges in the TDL~\cite{shepherd_many-body_2013} due to
the contribution of low-energy excitations with vanishing momentum transfer.
Because the long-range part of the Coulomb interaction is already treated with CCSD
(which treats density fluctuations at the level of the random-phase approximation~\cite{bohm_collective_1951,pines_collective_1952,bohm_collective_1953,pines_collective_1953,scuseria_ground_2008}), 
we only consider the CCSD(T)
correlation energy due to the short-range part of the Coulomb interaction. 
In practice,
we calculate the CCSD(T) correlation energy using a coarse $\vk$-point mesh,
which can be understood
as an approximate regularization of an infrared divergence via enforcement 
of a minimum momentum
transfer determined by the employed $\vk$-point mesh;
we will refer to this method as CCSD(T)$_\mathrm{SR}$.
Here we use $N_\vk=2^3$ for Li and $N_\vk=2\times 2\times 1$ for Al, which
corresponds to neglecting momentum transfers that are less than about
1~\AA$^{-1}$.
The EOS using this approach is shown in the middle panels of Fig.~\ref{fig:binding},
with properties given in Tab.~\ref{tab:properties}. The error in the
cohesive energy is reduced by about 30\%. Other properties are also improved,
but less significantly.

In our
second approximate improvement, which we refer to as CCSD with scaling all correlation 
(CCSD-SAC)~\cite{gordon_scaling_1986},
we calculate the correlation energy of the solid as
$E_\mathrm{c}^\textrm{CCSD-SAC} = E_\mathrm{c}^\textrm{CCSD}/F(\rho)$,
where $F(\rho)$ is a non-empirical factor that is determined according to the UEG
of the same density $\rho$ as the sytem under study.
Specifically, we take
$F(\rho) = E_\mathrm{c}^\mathrm{CCSD}(\rho)/E_\mathrm{c}^\mathrm{exact}(\rho)$, 
where $E_\mathrm{c}^\mathrm{exact}$
is the exact correlation energy of the UEG according to the Perdew-Zunger
fit~\cite{perdew_self-interaction_1981} 
to diffusion Monte Carlo results~\cite{ceperley_ground_1980} and
$E_\mathrm{c}^\mathrm{CCSD}$ is the CCSD correlation energy of the UEG according
to our own Perdew-Zunger-style fit to CCSD results from
Shepherd~\cite{shepherd_communication_2016}.
For this UEG description, we consider one valence electron for Li and three
valence electrons for Al, and calculate the density at all lattice parameters
studied. 
At the experimental lattice parameters,
this approach gives densities
corresponding to Wigner-Seitz radii of about $r_s = 3.24$ Bohr for Li and $r_s = 2.06$ Bohr for Al
and scaling factors $F \approx 0.8$--$0.9$.
The EOS calculated using CCSD-SAC is shown
in in the bottom row of Fig.~\ref{fig:binding}, with properties given in
Tab.~\ref{tab:properties}. The cohesive energies are significantly improved over
CCSD or CCSD(T)$_\mathrm{SR}$. For Li, the magnitude of the cohesive energy is overestimated by
4~m$E_h$ (7\% error), although structural properties are notably worse than those from
CCSD or CCSD(T)$_\mathrm{SR}$. For Al, the cohesive energy is almost perfectly predicted,
but structural properties are marginally improved (bulk modulus) or
slightly worse (lattice parameter).

\textit{Discussion and conclusions.}
For Li, our results can be compared to previous ones in the literature, which are included
in Tab.~\ref{tab:properties}. 
First, our HF results are in good agreement with those reported in Ref.~\onlinecite{paulus_hartreefock_2007},
which were calculated using an optimized DZ basis with the CRYSTAL package~\cite{saunders_crystal2003_2004}.
This agreement confirms a consistent starting point for correlated calculations.
To our knowledge, there are no reports of periodic CCSD calculations of the
equation of state of BCC Li (a recent work~\cite{mihm_effective_2021} 
used periodic CCSD to estimate the
energy difference between FCC and BCC Li).
However, in Ref.~\onlinecite{stoll2009}, an incremental CCSD scheme was applied
based on finite clusters.  Although that work found a lattice constant in good
agreement with ours, it found a cohesive energy that was significantly different
($-0.060$~$E_h$ compared to our $-0.051$~$E_h$) and in much better agreement
with experiment. As discussed in detail in that work, the application of
incremental schemes to metallic systems is delicate and nontrivial, and this 
might be responsible for the disagreement. Based on our investigations, we find
no evidence for errors on the order of 0.01~$E_h$. Instead, we believe that this
level of accuracy is expected for CCSD based on its known performance for the
UEG~\cite{callahan_dynamical_2021,shepherd_communication_2016}, where it
underestimates the correlation energy at metallic densities by about 10-20\%.

Finally, it is natural to compare CC methods to DFT, which is significantly more
affordable.  In Fig.~\ref{fig:strucprop}, we compare the properties of Li and Al
predicted by CCSD, CCSD(T)$_\mathrm{SR}$, and CCSD-SAC to those predicted by HF
and by common functionals, as reported in
Ref.~\onlinecite{zhang_performance_2018}. 
We compare to a few popular functionals of increasing
sophistication, including the local density approximation
(LDA)~\cite{kohn_self-consistent_1965}, the generalized gradient approximations
(GGAs) PBE~\cite{perdew_generalized_1996} and PBEsol~\cite{perdew_restoring_2008},
the meta GGA M06-L~\cite{zhao_new_2006}, and the screened hybrid
HSE06~\cite{heyd_hybrid_2003,heyd_erratum_2006,krukau_influence_2006}. Overall,
the CC results are comparable to those from GGAs but worse than the hybrid
HSE06, which performs extremely well for these two materials.

Looking forward, it will be interesting to apply CC methods to less uniform
metallic systems, such as metal surfaces including adsorbates or chemical
reactants. For these problems, we expect to see an increased advantage of CC
over DFT, due to the greater variations in the electron density and the
importance of dispersion interactions. More broadly, our work has reemphasized
the need for wavefunction based methods that improve upon CCSD without the use
of perturbation theories that diverge for metals. Beyond more systematic
investigation of the two methods proposed here, other possibilities include the
use of spin-component scaling~\cite{grimme_improved_2003,
takatani_improvement_2008, grimme_spin-component-scaled_2012},
regularization~\cite{lee_regularized_2018,shee_regularized_2021,keller_regularized_2022},
or screened
interactions~\cite{landsberg_contribution_1949,shepherd_many-body_2013},
although these approaches typically introduce empirical parameters.
Alternatively, the full or limited inclusion of non-perturbative
triple excitations~\cite{lee_coupled_1984,urban_towards_1985,shavitt_many-body_2009} 
is a promising
\textit{ab initio} route towards chemical accuracy in metallic solids.\\

\textit{Data Availability.} After acceptance, data will be made openly
available. To ask for further data, a reasonable request can be sent to the
authors.\\

\textit{Acknowledgments.} This work was primarily supported by the Columbia
Center for Computational Electrochemistry (V.A.N.) and partially supported by
the National Science Foundation under Grant Numbers OAC-1931321 (H.-Z.Y.) and
CHE-1848369 (T.C.B.).  We acknowledge computing resources from Columbia
University's Shared Research Computing Facility project, which is supported by
NIH Research Facility Improvement Grant 1G20RR030893-01, and associated funds
from the New York State Empire State Development, Division of Science Technology
and Innovation (NYSTAR) Contract C090171, both awarded April 15, 2010. Some
calculations were performed using resources provided by the Flatiron Institute.
The Flatiron Institute is a division of the Simons Foundation.
Matplotlib~\cite{hunter_matplotlib_2007}, NumPy~\cite{harris_array_2020},
pandas~\cite{mckinney_data_2009}, SciPy~\cite{scipy_2020},
seaborn~\cite{Wseaborn2021}, and ColorBrewer~\cite{brewer_colorbrewer2_nodate} 
were used for calculations, analysis, and visualization.

\end{document}